\documentclass[aps, prb, amsfonts, amsmath, twocolumn, superscriptaddress, 10pt, floatfix]{revtex4-2}

\usepackage{graphicx}% Include figure files
\usepackage{bm}% bold math
\usepackage{subfig}
\usepackage{float}
\usepackage{comment}
\usepackage{lipsum} % for filler text
\usepackage{xcolor}
\usepackage{subcaption} %  for subfigures environments
\begin{document}

%\title{Optical Manipulation of Solitons Dynamics in Bose-Einstein Condensates within Dark Traps.}

\title{Optically Tuned Soliton Dynamics in Bose-Einstein Condensates within Dark Traps}

\author{Erwan C\'elanie}
 \affiliation{LyRIDS, ECE Paris, Graduate School of Engineering, Paris, F-75015, France} 
\author{Laurent Delisle}
 \affiliation{LyRIDS, ECE Paris, Graduate School of Engineering, Paris, F-75015, France} 
\author{Amine Jaouadi}
\email{ajaouadi@ece.fr}
\affiliation{LyRIDS, ECE Paris, Graduate School of Engineering, Paris, F-75015, France}

\date{\today}% It is always \today, today,

\begin{abstract}
This study investigates the formation and dynamics of solitons in Bose-Einstein condensates (BECs) within dark traps generated by two crossed Laguerre-Gaussian (LG) beams with varying azimuthal indices $\ell$. As the index $\ell$ increases, the potential transitions from a harmonic trap when $\ell = 1$ to a square-well potential for larger values of $\ell$. This transition allows us to study a range of soliton dynamics under different confinement conditions while maintaining the same BEC volume.
Through the derivation of the Gross-Pitaevskii equation (GPE) and under these specific conditions in both one-dimensional (1D) and two-dimensional (2D) configurations, we explore the dynamics of solitons across multiple scenarios. The study examines two primary methods for solitons generation: the temporal modulation of the scattering length and the implementation of an initial potential barrier that is subsequently removed. The results indicate that the trap shape plays a critical role in the generation and interaction dynamics of solitons. In harmonic traps, solitons exhibit a behavior different from those observed in anharmonic traps, where the dynamics is significantly influenced by the azimuthal index of the trap. The ability to control soliton dynamics in BECs holds significant promise for applications in quantum technologies, precision sensing, and the exploration of fundamental quantum phenomena.
\end{abstract}
\maketitle
%******************************
\section{Introduction}
%******************************
Quantum theory was primarily developed to describe microscopic phenomena such as the structure of the atom and chemical bonds. Nevertheless, quantum mechanics can manifest on macroscopic scales, with superconductivity and superfluidity being prime examples. It took more than 70 years to experimentally verify the theoretical predictions of Bose-Einstein condensation. In 1995, Eric Cornell, Carl Wieman, and Wolfgang Ketterle observed this remarkable phenomenon in an ultra-cold dilute gas of $^{87}R$b atoms \cite{Anderson1995}. This groundbreaking achievement earned them the Nobel Prize in Physics in 2001 \cite{nobel2001}. With this experimental confirmation of BEC condensation, quantum degeneracy has become accessible for a wide variety of atomic species. Researchers have successfully created BECs with elements such as alkali metals \cite{Anderson1995, Weber2003, Davis1995, Bradley1995, Roati2007}, hydrogen \cite{Fried1998}, metastable helium \cite{Robert2001, Santos2001}, ytterbium \cite{Fukuhara2007}, and chromium \cite{Beaufils2008, Griesmaier2005}. This discovery has facilitated the development of a highly active field of research. The number of research groups studying ultra-cold diluted gases has grown exponentially. The scientific community's interest in BECs is driven by their intersection with various fields of physics, including condensed matter, atomic physics, and quantum optics.
In condensed matter systems, interactions between particles often obscure quantum statistical effects. In contrast, an ultra-cold gas such as a BEC offers a perfect environment for studying the collective behavior of the particles composing it. These particles can evolve in periodic or disordered optical structures, allowing for the exploration of quantum phenomena with unprecedented clarity and control. The ability to manipulate and control BECs with optical and magnetic fields has opened new avenues for research \cite{Gaaloul2007,Gaaloul2009,Jaouadi2010, Mewes1997}. Scientists are now exploring the use of BECs in creating novel quantum devices, such as atom lasers \cite{Bloch1999} and quantum sensors \cite{MORALES2022}. Additionally, the generation and control of solitons in BECs have emerged as a significant area of interest \cite{khaykovich2002, denschlag2000, eiermann2004}. In fact, solitons manifest themselves, in different domains, as stable localized waves that maintain their shape while propagating \cite{kosevich1990magnetic, wu1984observation, mollenauer1980, tanaka2001soliton}. Their existence in BECs provides valuable insight into nonlinear dynamics and quantum coherence \cite{Alison2022}. 
These solitary waves can be used in precision measurement \cite{Yang23}, information processing, and studying fundamental quantum phenomena. The rich interplay of theory and experiment in this field continues to push the boundaries of the understanding of quantum mechanics and its applications. 

Solitons in BECs have been extensively explored within the framework of harmonic traps \cite{David2012,Wang,Henderson2020,Cornish2006,Nguyen2014,Strecker2002,Carretero-Gonzalez2008}. These traps, in the form of a quadratic potential, allow the creation and observation of both bright and dark solitons \cite{David2012,2019_Hussain} depending on the process used. The harmonic trapping potential offers a well-controlled environment in which the interactions and stability of solitons can be mathematically investigated. In this study, we investigate the generation and dynamics of solitons within BECs subjected to shaped potentials created by Laguerre-Gaussian beams as presented in \cite{Jaouadi2010}. LG beams are distinguished by their phase structure and the ability to carry orbital angular momentum, leading to tailored potential shapes that can profoundly influence soliton behavior \cite{allen1992orbital}. By manipulating parameters such as the beam waist and the orbital angular momentum, we can tailor these potentials, providing control over soliton dynamics.
Understanding the effects of these shaped potentials on solitons is essential to explore soliton behavior and interaction. The stability and robustness of solitons in BECs make them ideal candidates for applications in quantum information processing, precision measurements, and quantum sensing. In addition, solitons occurring in BECs can offer new perspectives for studying nonlinear phenomena in quantum systems. They can serve as carriers of quantum information, with their interactions potentially useful for creating entangled states and performing quantum logic operations.

In this paper, we present both theoretical and numerical analysis of soliton solutions in BECs, employing two crossed blue-detuned LG beams to form a 3D dark trap, where the BEC will be formed. LG beams create a tunable trapping potential from harmonic potential to a cubic potential by changing the azimuthal index $\ell$. Within this framework, we explore the dynamic of solitons in both 1D and 2D configurations. The findings of this research extend our understanding of soliton dynamics in non-traditional potentials and emphasize the practical implications for advanced quantum technologies.

The outline of the paper is as the following: In Section II, we present the theoretical model and the creation of the 3D traps using LG beams. Section III discusses the dimensional reductions of the 3D Gross-Pitaevskii (GP) equation describing the dynamics in 1D and 2D configurations. In Section IV, we present the numerical results and discussions in both 1D and 2D settings after presenting two methods for generating solitons. In Section V, we conclude with a summary of our findings and present some future perspectives.

%****************************
\section{The Model}
\begin{figure}[h]
    \centering
    \includegraphics[width=\linewidth]{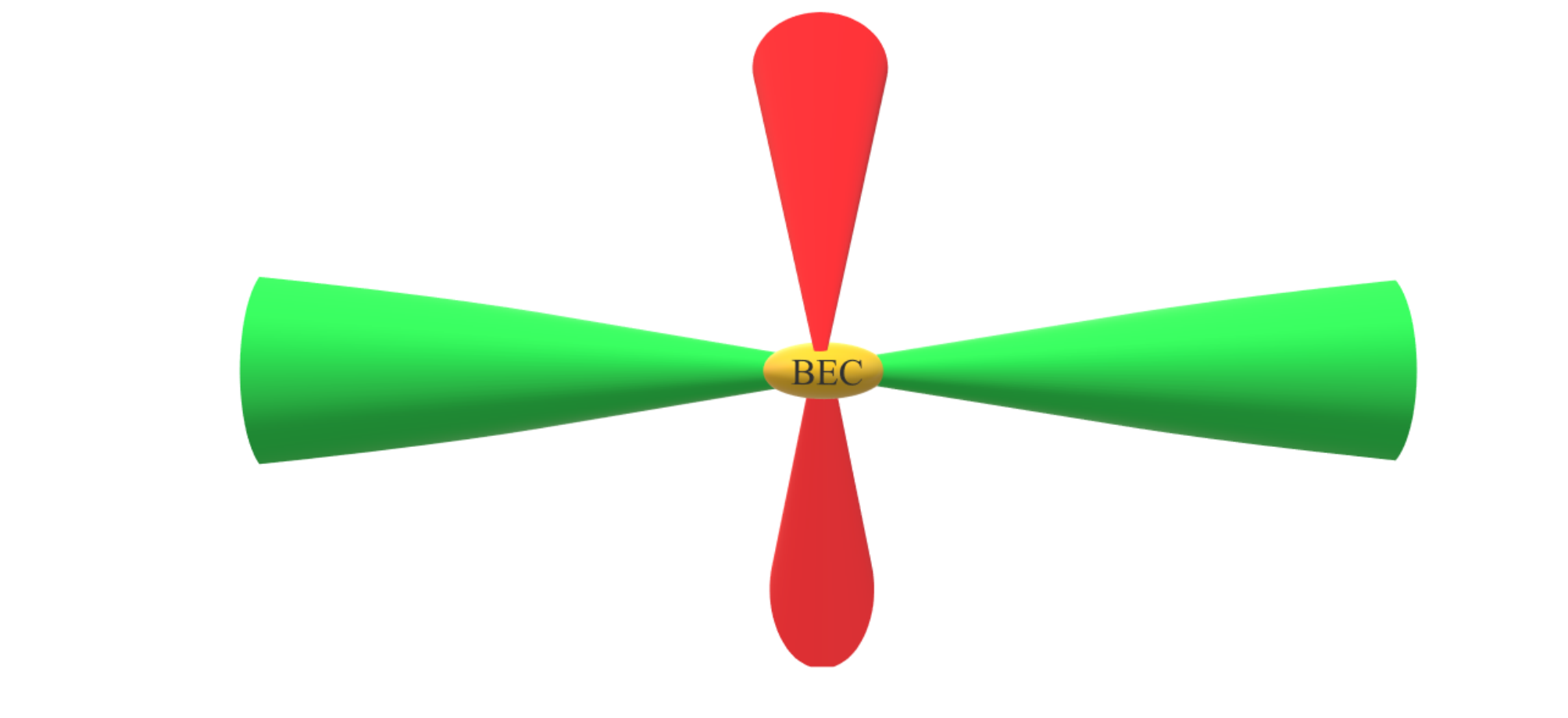}
    \caption{The BEC is formed at the intersection of the two LG beams. }
    \label{fig:Fig1}
\end{figure}

%****************************
In our model, we aim to confine atoms in an "all-optical" trap with the objective of achieving BECs. We consider here various elements such as Rubidium ($^{87}$Rb), Sodium ($^{23}$Na) and Lithium ($^7$Li). The proposed experiment, presented initially in \cite{Jaouadi2010}, involves creating a three-dimensional trap formed by two crossed Laguerre-Gaussian laser beams, with the first beam propagating along the z-direction and the second beam propagating in a perpendicular direction as presented in figure \ref{fig:Fig1}.
%************************************
\subsection{Laguerre-Gauss Beams}
\begin{figure}[h]
    \centering
    \includegraphics[width = 6 cm, height = 6 cm ]{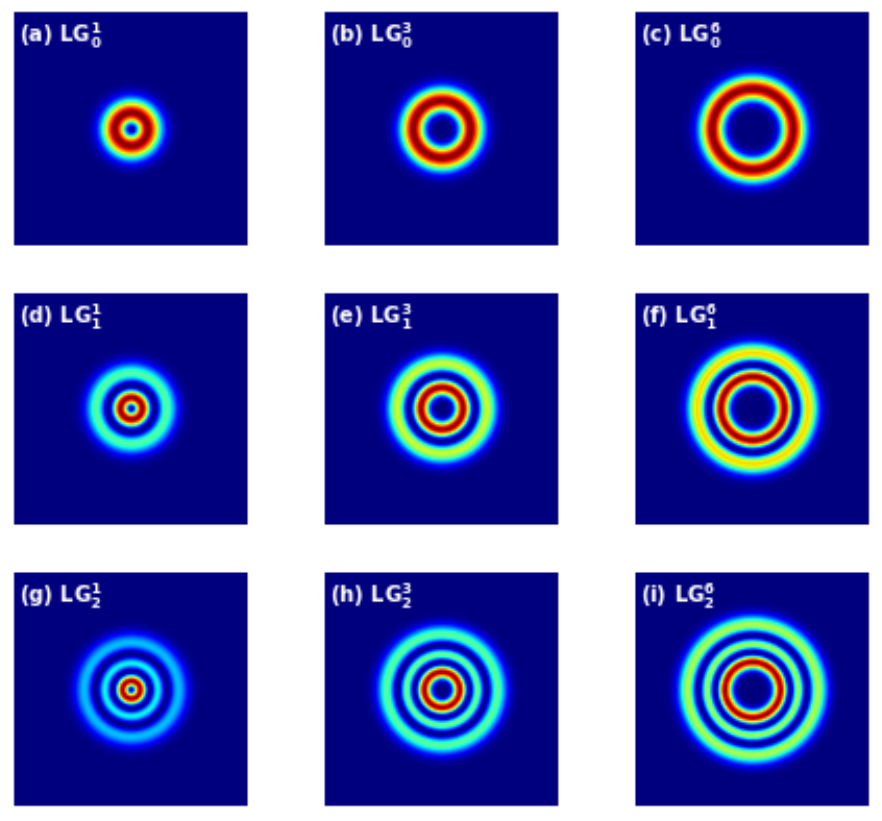}
    \caption{Intensity profiles of Laguerre-Gaussian beams \( L_p^{|\ell|} \) for different values of $\ell$ and $p$.}
    \label{fig:Fig2}
\end{figure}

%************************************
Laguerre-Gaussian (LG) beams are a class of laser modes that exhibit unique phase and intensity characteristics, making them valuable tools in a variety of optical and quantum applications. These beams are described by their azimuthal phase dependence \( \exp(i\ell\phi) \), where \(\ell\) denotes the orbital angular momentum (OAM) quantum number. The parameter \(\ell\) signifies the number of twists the phase front of the beam undergoes as it propagates, imparting a helical structure to the wavefront. This helical wavefront results in the LG beam carrying orbital angular momentum, a property that distinguishes it from other beam types.
The intensity profile of an LG beam presented in figure \ref{fig:Fig2}, characterized by mode indices \(\ell\) and \(p\), is given by the following expression:
\begin{equation}
\begin{split}
I_{\ell p}(\rho, z) & = \frac{P_0}{\pi w(z)^2}\left(\frac{\rho\sqrt{2}}{w(z)}\right)^{2|\ell|} L_p^{|\ell|} \left( \frac{2\rho^2}{w(z)^2} \right) \\
& \times \exp \left( - \frac{2\rho^2}{w(z)^2} \right),
\end{split}
\end{equation}
where \( \rho \) is the radial distance from the beam axis, \( w(z) \) is the beam waist, which varies with the axial distance \(z\) according to the beam's divergence. \( L_p^{|\ell|} \) is the associated Laguerre polynomial, which determines the radial structure of the beam. The index \(p\) indicates the number of radial nodes in the intensity profile. \( P_0 \) is the total power of the beam.
The beam waist \(w(z)\) is a function of the propagation distance \(z\) and can be expressed as:
\begin{equation}
w(z) = w_0 \sqrt{1 + \left( \frac{z}{z_R} \right)^2},
\end{equation}
where \(w_0\) is the beam waist at the focus ($z = 0$), and \(z_R\) is the Rayleigh range, defined as: $z_R = \frac{\pi w_0^2}{\lambda}$ 
with \(\lambda\) being the wavelength of the laser light.
%*****************************************
\subsection{Shaped traps}
\begin{figure}[h]
    \centering
    \includegraphics[width=\linewidth, height = 4 cm]{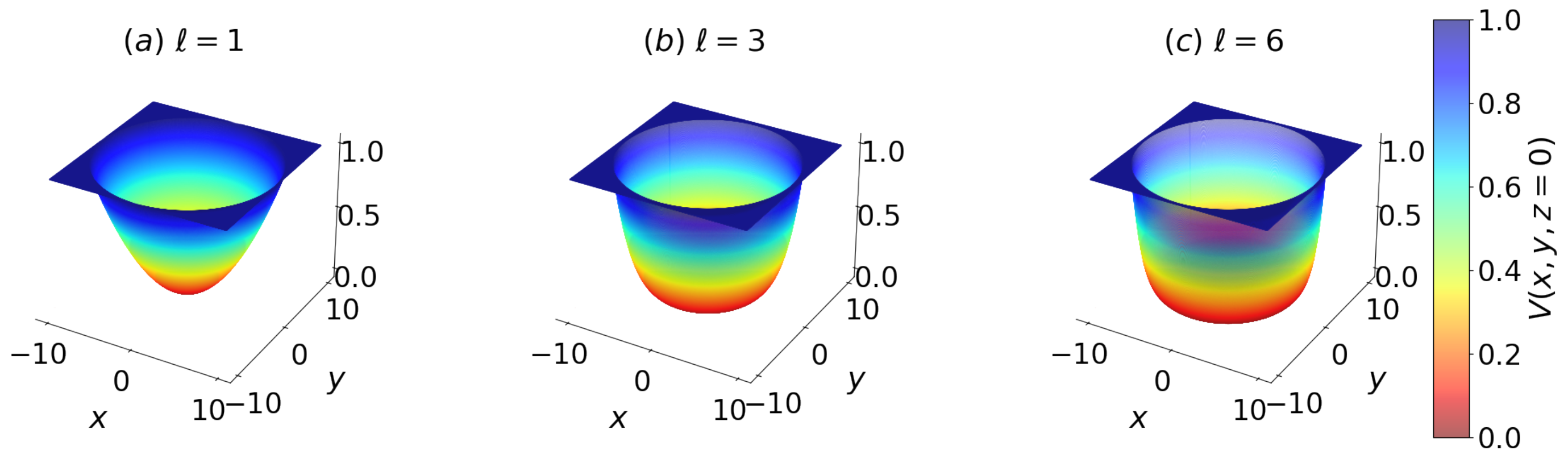}
    \caption{3D potential for $\ell=1,3,6$. }
    \label{fig:Fig3}
\end{figure}

%*****************************************
When two LG beams intersect orthogonally, the resulting potential is presented in figure \ref{fig:Fig3}. In the limit where $\rho = \sqrt{x^2 + y^2} << w_0$, we obtain a highly simplified form of the potential that closely resembles to a power-law potential as:
\begin{equation}
V(\rho, z) = U_{\rho}\rho^{2\ell} + U_zz^{2\ell},\label{PotentialV}
\end{equation}
where \( U_{\rho} \) and \( U_z \) are the potential depths along the respective axes. They are directly determined by the laser characteristics, including the power \( P_0 \), the laser detuning $\delta$, and the waist $w_0$  of the two beams \cite{Jaouadi2010}.

%***********************************************
\section{Dimension reduction}
%***********************************************

The dynamics of the BEC at zero temperature are governed by the Gross-Pitaevskii equation (GPE), a nonlinear Schrödinger equation that describes the macroscopic wave function of the condensate. The GPE is given by:
\begin{equation}
i\hbar\frac{\partial \psi(\mathbf{r}, t)}{\partial t} = \left( -\frac{\hbar^2}{2m}\nabla^2 + V(\mathbf{r}, t) + Ng|\psi(\mathbf{r}, t)|^2 \right) \psi(\mathbf{r}, t).\label{GPeq}
\end{equation}

In this equation, \( \psi(\mathbf{r}, t) \) represents the condensate wave function. The term \( \frac{\hbar^2}{2m}\nabla^2 \) corresponds to the kinetic energy of the particles, where \( \hbar \) is the reduced Planck constant, and \( m \) is the atomic mass. The external potential \( V(\mathbf{r}, t) \) can vary spatially and temporally, influencing the behavior and dynamics of the condensate. The interaction term \( Ng|\psi(\mathbf{r}, t)|^2 \) accounts for the mean-field interactions between the particles in the condensate, with \( N \) being the number of particles and \( g = 4\pi\hbar^2a_s/m \) representing the interaction strength characterized by the length of s-wave scattering \( a_s \).

To explore the behavior of solitons in Bose-Einstein Condensates, we investigate soliton solutions in 1D and 2D spatial dimensional spaces. This involves starting from the 3D GPE (\ref{GPeq}) and employing dimensional reduction techniques to derive a corresponding 1D or 2D equation accordingly. The process of dimensional reduction is typically justified when the system exhibits strong confinement in one spatial dimension, such that the dynamics along this dimension is effectively ``frozen'' and the system can be described by a lower-dimensional equation.

%************************************
\subsection{1D-BEC reduction}
%************************************

For the 1D reduction, we assume that $\frac{U_{\rho}}{{U_z}}<\!\!< 1$ in the trap potential $V$ given in (\ref{PotentialV}). This assumption implies that the motion of atoms in the $(x,y)$ plane direction is frozen into the ground state, allowing the system to be treated as one-dimensional.

Let us assume that the wave function is composed of a constant Gaussian transverse part $\phi_{0}$, which is the ground state of the transverse harmonic potential, and a time-varying axial component $f$ as: 
\begin{equation}
\psi(\mathbf{r}, t)=f(z, t) \phi_{0}(x) \phi_0(y), \quad \phi_0(\zeta)=\frac{1}{\sqrt{2\pi}\sigma}e^{-\frac{\zeta^2}{2\sigma^2}},  \label{ansatz1D}
\end{equation}
where $\zeta = x, y$. The parameter $\sigma$ will be chosen in future steps. The Gaussian function $\phi_0$ satisfies
$$\phi_{0,\zeta\zeta}(\zeta)=\frac{1}{\sigma^4}(\zeta^2-\sigma^2)\phi_0(\zeta),$$ 
which reveals its importance in calculations. For the 1D reduction of the GP equation (\ref{GPeq}), we suppose the following form for the potential $V$:
$$V(\mathbf{r},t)=V(\mathbf{r})=\frac12 m\omega_{\perp}^2(x^2+y^2)+V_{1D}(z),$$
where $V_{1D}$ is an arbitrary function of the variable $z$. Introducing the ansatz (\ref{ansatz1D}) into the 3D GP equation (\ref{GPeq}) yields the following equation:
\begin{equation}
    i\hbar f_t=-\frac{\hbar^2}{2m}f_{zz}+\hbar \omega_{\perp} f+V_{1D} f+Ng\vert\phi_0(x)\vert^2\vert\phi_0(y)\vert^2\vert f\vert^2f,\label{1DGPpart1}
\end{equation}
where $\sigma^2=\frac{\hbar}{m\omega_{\perp}}$. The term $\hbar \omega_{\perp}f$ can be ignored using the following rotational transformation:
$$f(z,t)=e^{i\mu t}F(z,t),\quad \mu=-\omega_{\perp}.$$
Indeed, introducing this transformation into equation (\ref{1DGPpart1}) and integrating over the variables $x$ and $y$ yields to the desired equation:
\begin{equation}
    i\hbar F_t=-\frac{\hbar^2}{2m}F_{zz}+V_{1D} F+Ng\eta \vert F\vert^2 F \label{1DGPfinal},
\end{equation}
where
$$\eta=\frac{\int_{\mathbb{R}^2}\vert\phi_0(x)\vert^4\vert\phi_0(y)\vert^4\mbox{d}x\mbox{d}y}{\int_{\mathbb{R}^2}\vert\phi_0(x)\vert^2\vert\phi_0(y)\vert^2\mbox{d}x\mbox{d}y}=\frac{m^2\omega_{\perp}^2}{8\pi^2\hbar^2}.$$

Hence, to resume, the transformation
$$\psi(\mathbf{r},t)=\frac{m\omega_{\perp}}{2\pi\hbar}F(z,t)\times\exp\left(-\frac{\omega_{\perp}}{2\hbar}(m\rho^2+2i\hbar t)\right)$$
allows one to reduce the 3D GP equation (\ref{GPeq}) into the 1D GP equation (\ref{1DGPfinal}).
In this study, the potential $V_{1D}$ will be chosen as
$$V_{1D}(z)=U_zz^{2l}.$$
An analysis will be performed following different values of the integer $\ell$. In particular for $\ell=1$, the potential is said to be harmonic and anharmonic for other values.

%**************************************
\subsection{2D-BEC reduction}
%**************************************

For the 2D reduction, we assume that $\frac{U_z}{U_{\rho}}<\!\!< 1$ in the potential $V$ given in (\ref{PotentialV}). This assumption implies that the motion of atoms in the $z$ direction is frozen into the ground state and that the system can be viewed as two-dimensional.

Let us assume that the wave function is composed of a constant Gaussian axial part $\phi_{0}$, which is the ground state of the axial harmonic potential, and a time-varying transverse component $f$ as
\begin{equation}
\psi(\mathbf{r}, t)=f(x,y, t) \phi_{0}(z),\quad \phi_0(z)=\frac{1}{\sqrt{2\pi}\sigma}e^{-\frac{z^2}{2\sigma^2}},\label{ansatz2D}
\end{equation}
where $\sigma$ will be carefully chosen in the following dimensional reduction steps. For this reduction and the aims of this paper, we choose the potential $V$ in equation (\ref{GPeq}) as
$$V(\mathbf{r},t)=V(\mathbf{r})=V_{2D}(\rho)+\frac12 m\omega_z^2z^2,$$
where $V_{2D}$ is an arbitrary function of the radial variable $\rho$.

Introducing ansatz (\ref{ansatz2D}) into equation (\ref{GPeq}) yields the following equation for function $f$:
\begin{equation}
i\hbar f_t=-\frac{\hbar^2}{2m}\nabla^2_{2D}f+\frac{\hbar\omega_z}{2}f+V_{2D}f+Ng\vert \phi_0(z)\vert^2\vert f\vert^2f,\label{2Dred1}
\end{equation}
for $\sigma^2=\frac{\hbar}{m\omega_z}$. The term $\frac{\hbar\omega_z}{2}f$ can be ignored using the following rotational transformation:
$$f(x,y,t)=e^{i\mu t}F(x,y,t),\quad \mu=-\frac{\omega_z}{2}.$$
Indeed, introducing this transformation into equation (\ref{2Dred1}) and integrating over the variable $z$, yields the desired equation:
\begin{equation}
i\hbar F_t=-\frac{\hbar^2}{2m}\nabla_{2D}^2F+V_{2D}F+Ng\eta \vert F\vert^2F,\label{2DredFinal}
\end{equation}
where 
$$\eta=\frac{\int_{\mathbb{R}}\vert\phi_0(z)\vert^4\,\mbox{d}z}{\int_{\mathbb{R}}\vert\phi_0(z)\vert^2\,\mbox{d}z}=\frac{\sqrt{2}m\omega_z}{4\pi\hbar}.$$
Hence, to resume, the transformation
$$\psi(\mathbf{r},t)=\sqrt{\frac{m\omega_z}{2\pi\hbar}}F(x,y,t)\times \exp\left(-\frac{\omega_z}{2\hbar}(mz^2+i\hbar t)\right)$$
allows one to reduce the 3D GP equation (\ref{GPeq}) into the 2D GP equation (\ref{2DredFinal}).
In the context of our paper, the potential $V_{2D}$ will be chosen as
$$V_{2D}(\rho)=U_{\rho}\rho^{2l}.$$
An analysis will be realized following different values of the integer $l$. In particular for $l=1$, the potential is said to be harmonic and anharmonic for other values.

\subsubsection{Hirota bilinear formalism for $\ell=1$}

In order to find exact soliton solutions of equation (\ref{2DredFinal}) with the harmonic potential $V_{2D}=U_{\rho}\rho^{2}=\frac{1}{2}m\omega_{\perp}^2 \rho^{2}$, one can use the so-called Hirota bilinear method \cite{Hirota}. This algebraic method allows one to cast a nonlinear partial differential equation into a bilinear equation from which multisoliton solution can be obtained from a sum of exponential functions.

Let us consider the following transformation for the function $F$ in equation (\ref{2DredFinal}): 
\begin{equation}
F(x,y,t)=H(X,Y,T) \times \exp\left(r+i q (X^2+Y^2)\right),
\label{ansatzHirota}
\end{equation}
where $r=r(T)$ and
$$X=e^{r}x,\quad Y=e^{r}y,\quad T=t,\quad q=-\frac{m}{2\hbar}e^{-2r}\dot{r}.$$

Using the ansatz (\ref{ansatzHirota}) and choosing $r$ to satisfy the differential equation
$$\ddot{r}=(\dot{r})^2+\omega_{\perp}^2\quad \mbox{with}\quad\dot{r}=\frac{\mbox{d}r}{\mbox{d}T}$$
allows one to transform equation (\ref{2DredFinal}) into a Schrödinger-type equation:
\begin{equation}
    i\hbar H_T= e^{2r}\left(-\frac{\hbar^2}{2m}\nabla_{2D}^2H+Ng\eta\vert H\vert^2H\right).\label{Schro}
\end{equation}
The differential equation satisfied by $r$ can be integrated to obtain the following exact expression:
$$r(T)=-\ln \vert\cos(\omega_{\perp}T+\alpha_1)\vert+\alpha_2,$$
where $\alpha_1$ and $\alpha_2$ are constants of integration. The above observations suggest that equation (\ref{Schro}) is invariant under time translation $T\longrightarrow T+\frac{2\pi}{\omega_{\perp}}$, meaning that, as $\omega_{\perp}$ increases, the period decreases, implying more collisions between solitons. This manifestation can be seen, in the 1D spatial dimensional space, in figure \ref{fig:Fig4}.\\
Equation (\ref{Schro}) as been largely studied and have shown to possess multisoliton solutions. These solutions as been studied and constructed in \cite{Wang}. They constructed these solutions using an Hirota bilinear transformation \cite{Hirota}. Indeed, taking
$$H=\frac{C}{R},$$
 where $C$ and $R$ are, respectively, complex and real valued functions allows one to transform equation (\ref{Schro}) into bilinear form:
 \begin{eqnarray}
\left(i \mathcal{D}_T+\frac{\hbar}{2m}e^{2r}(\mathcal{D}_X^2+\mathcal{D}_Y^2)\right)(C\cdot R)=0,\label{bilinear1}\\
\frac{\hbar^2}{2m}(\mathcal{D}_X^2+\mathcal{D}_Y^2)(R\cdot R)+Ng\eta \vert C\vert^2=0.\label{bilinear2}
 \end{eqnarray}
 The operator $\mathcal{D}$ is known as the Hirota bilinear derivative and is defined as
 $$\mathcal{D}_{\mu}^n(F_1\cdot F_2)=(\partial_{\mu_1}-\partial_{\mu_2})^n F_1(\mu_1)F_2(\mu_2)\vert_{\mu_1=\mu_2=\mu}.$$
 Below, we construct the travelling one-soliton solution as an example and we refer the reader to \cite{Wang} for the construction of multisoliton solutions.\\
 One should note that an open problem is to construct multisoliton solutions in the case of $\ell\neq 1$ for the potential $V_{2D}=U_{\rho}\rho^{2\ell}$. This investigation is part of a future research project.
 %****************************************************
\subsubsection{Exact soliton solution for $\ell=1$}
%*****************************************************
In order to construct multisoliton solutions, we use perturbation theory around the trivial solution $C=0$ and $R=1$ in the bilinear equations (\ref{bilinear1}) and (\ref{bilinear2}). For the travelling wave solution (or one-soliton solution), we suppose
$$C=\epsilon C_1\quad \mbox{and}\quad R=1+\epsilon^2 R_2,$$
where $\epsilon$ is a free real parameter, $C_1$ is a complex-valued function and $R_2$ a real-valued function. Introducing these forms into the bilinear equations allows one to get, for each powers of $\epsilon$, this system:
\begin{eqnarray}
    i\partial_TC_1+\frac{\hbar}{2m}e^{2r}\nabla_{2D}^2C_1=0,\label{T1}\\
    \left(i \mathcal{D}_T+\frac{\hbar}{2m}e^{2r}(\mathcal{D}_X^2+\mathcal{D}_Y^2)\right)(C_1\cdot R_2)=0,\label{T2}\\
    \frac{\hbar^2}{m}\nabla_{2D}^2R_2+Ng\eta \vert C_1\vert^2=0,\label{T3}\\
    (\mathcal{D}_X^2+\mathcal{D}_Y^2)(R_2\cdot R_2)=0\label{T4}.
\end{eqnarray}
For the travelling wave solution, we suppose that
$$C_1=e^{\Lambda}, \quad \Lambda=\kappa_X X+\kappa_Y Y+\lambda(T),$$
where $\kappa_X$, $\kappa_Y$ are constants and $\lambda$ is a pure imaginary function of $T$. Introducing this form into equation (\ref{T1}) yields the following differential equation for $\lambda$:
$$i\dot{\lambda}+\frac{\hbar}{2m}e^{2r}\left(\kappa_X^2+\kappa_Y^2\right)=0.$$
This equation can be integrated to get the exact form of $\lambda$:
$$\lambda(T)=i\frac{\hbar}{2m\omega_{\perp}}e^{2\alpha_2}\left(\kappa_X^2+\kappa_Y^2\right)\tan(\omega_{\perp} T+\alpha_1).$$
Introducing $C_1$ in equation (\ref{T3}) gives the form for the function $R_2$:
$$R_2=\xi e^{\Lambda+\Lambda^{*}},\quad \xi=-\frac{Ng\eta m}{4\hbar^2\left(\kappa_X^2+\kappa_Y^2\right)}.$$
The expressions for $C_1$ and $R_2$ insure that the system of equations (\ref{T1})-(\ref{T4}) is satisfied.

To resume, we get an exact solution $\psi$ where $\vert\psi\vert^2$ is given by
$$\vert\psi\vert^2=\frac{\epsilon^2m\omega_z}{2\pi\hbar}\times\frac{\exp\left(\Lambda+\Lambda^{*}+2r-\frac{\omega_zm}{\hbar}z^2\right)}{\left(1+\epsilon\xi\exp(\Lambda+\Lambda^*)\right)^2}.$$

For interactions between $N\in\mathbb{N}^*$ solitons, the exact expressions are obtained following a similar procedure. Indeed, in this case, one suppose that the functions $C$ and $R$ take the forms:
$$C=\sum_{k=1}^N\epsilon^{2k-1}C_{2k-1}\quad \mbox{and}\quad R=1+\sum_{k=1}^N\epsilon^{2k}R_{2k},$$
where the functions $C_j$ and $R_j$ are, respectively, complex-valued and real-valued. Introducing these forms into equations (\ref{bilinear1}) and (\ref{bilinear2}) yields a system of equations for each powers of $\epsilon$. These equations can then be solved using a sum of independent exponential functions.

%*****************************
\section{Soliton Dynamics}
%*****************************
There are several methods to generate bright or dark solitons in BECs \cite{Inouye1998, Chin2010, Burger1999, Anderson2001, Strecker2002, denschlag2000, khaykovich2002}. In this study, we focus on two distinct methods:
\begin{enumerate}
    \item The first method involves dynamically varying the scattering length \(a_s\), which controls the interaction between the atoms within the BEC. Initially, we take a positive value as \(a_s = 1.5 \, \text{nm}\) when \(t_s < 200 \, \text{ms}\). After this time, we switch to a negative value, such as \(a_s = -0.2 \, \text{nm}\). This abrupt change induces attractive interactions between the atoms, leading to the formation of solitons.
    This technique is widely used. For example, \textit{Strecker and al.} demonstrated the formation of solitons by dynamically tuning the length of the scattering using a Feshbach resonance \cite{David2012, Strecker2002}. The approach is effective because rapid changes in the strength of the interaction create localized stable solitons in the condensate.

    \item The second method for generating solitons involves the introduction of a potential barrier at the beginning of the experiment, which is subsequently removed. Initially, the BEC is trapped within the trap potential but with an additional barrier. By carefully controlling the height and width of this barrier, the condensate can be split into different parts. Once the desired initial state is achieved, the barrier is suddenly removed. This release allows the BEC to evolve naturally, forming solitons as a result of the internal dynamics and interaction of the BEC. This technique has been explored by various researchers, including the work by \textit{Cornish and al.} \cite{Cornish2006}, where solitons were generated by manipulating an additional barrier.
\end{enumerate}

We initialize the BEC wave function using the Thomas-Fermi approximation, where the chemical potential depends on the index $\ell$ of the trapping potential. The time-dependent Gross-Pitaevskii equation (\ref{GPeq}) is solved using the split-step Fourier method, alternating between position space for potential and interaction terms and momentum space for kinetic terms. We employ the imaginary-time relaxation technique \cite{Kosloff1986} to find the ground state, followed by real-time evolution to observe the dynamics of the solitons.
%****************************************
\subsection{SOLITON DYNAMICS IN 1D-BEC}
%*****************************************
%*********************************************
\subsubsection{Impact of Trap shape on Soliton Dynamics}
%*********************************************
%********************************
%\paragraph{Harmonic trap}
%********************************
%\hfill \break
%\hfill \break
To investigate how the shape of the potential affects the dynamics of the solitons, we first explore a simple 1D case. We examine the example of the harmonic potential with $\ell = 1$ as mentioned previously. For this study, we used $^7$Li atom as an example, keeping the mass fixed and varying only the depth of the potential. This approach allows us to have a straightforward 1D harmonic potential by altering its shape, which is achieved by changing the value of $U_z$ in $V_{1D}$. 
\begin{figure}[h]
    \centering
    \includegraphics[width=1\linewidth,height=5cm]{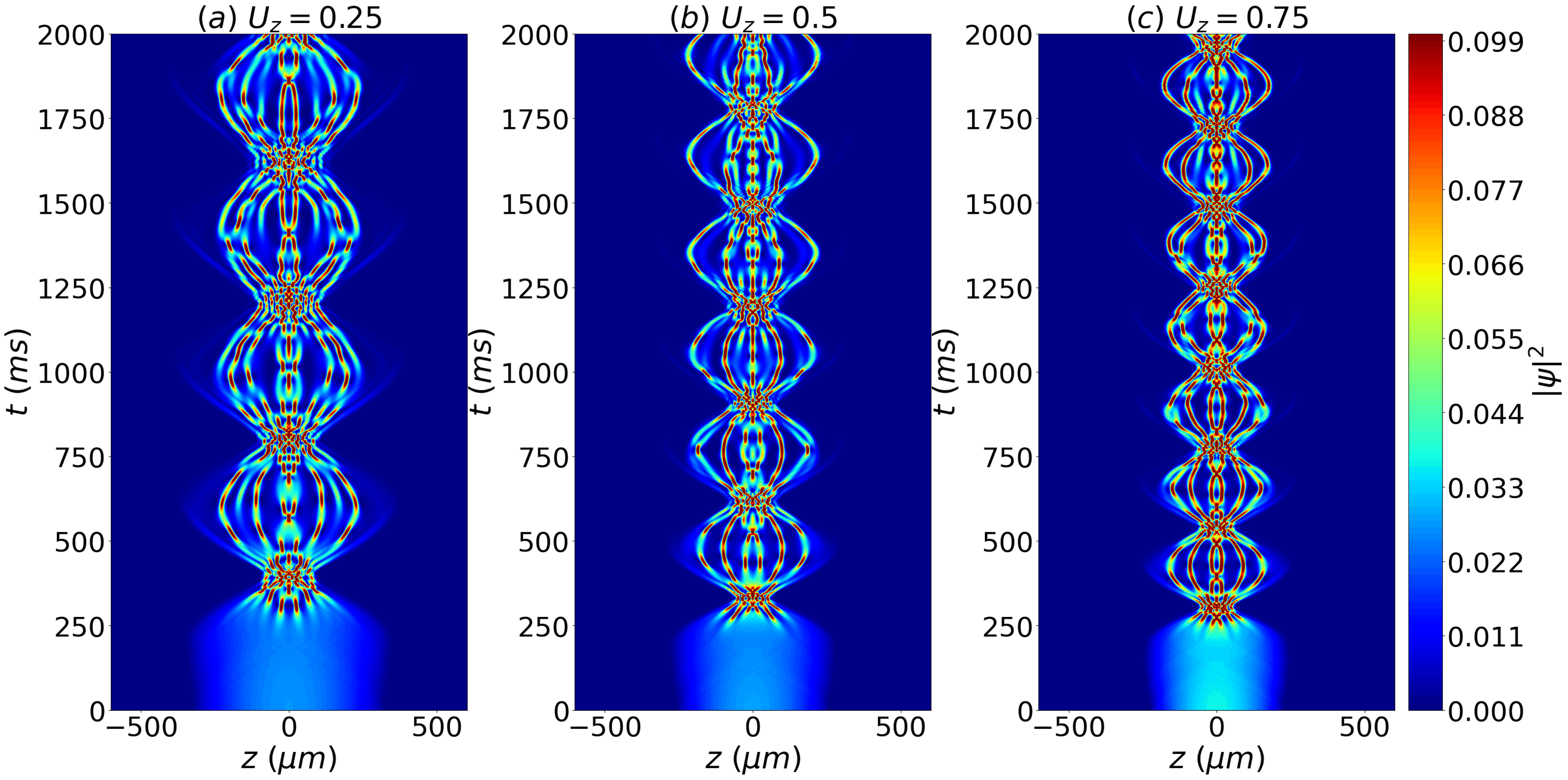}
    \caption{Soliton dynamics generated by changing the value of $a_s$ over time for different values of $U_z$ for $^7$Li atom.}
    \label{fig:Fig4}
\end{figure}
\hfill \break
\hfill \break
In figure \ref{fig:Fig4}, we present the soliton dynamics in the case of an harmonic trap ($\ell =1$) by adopting the first method in which we change the value of $a_s$ from a positive value to a negative one. The atom chosen here is the $^7$Li with a total number $N = 10^5$ of atoms. We choose \(a_s = 1.5 \, \text{nm}\) when \(t_s < 200 \, \text{ms}\) and, then, we switch to \(a_s = -0.2 \, \text{nm}\) at \(t_s < 200 \, \text{ms}\). 
The first notable observation, from figure \ref{fig:Fig4}, is the generation of what we call bright solitons, where maximum values of density are observed. In panel (a), which corresponds to $U_z$ = 0.25, 6 bright solitons are generated. These solitons begin to form around \( t_s = 250 \, \text{ms} \). The process initiates with the expansion of the BEC until \( t_s \), during which the interactions between the atoms are repulsive (\( a_s > 0 \)). The abrupt change in \( a_s \) from positive to negative values induces attractive interactions between the atoms, leading to the formation of solitons. After \( t_s \), the solitons propagate within the trap. The formation of solitons can be explained by the balance between dispersion and nonlinearity in the GP equation. When the scattering length is negative, attractive interactions overcome the dispersive effects, leading to the formation of localized soliton structures.
For \( U_z = 0.25 \) (panel (a)), the trap is relatively wide. The wider trap provides more space for solitons to propagate without frequent collisions, we observe in this case only 4 collisions. In contrast, as \( U_z \) increases, indicating a reduction in frequency along the \( z \) axis, the trap becomes progressively tighter. This increased tightness of the trap leads to a higher frequency of collisions between the solitons (refer to panels (b) and (c) of figure \ref{fig:Fig4}). The tight confinement enhances the interactions between the solitons as a result of the limited spatial region available for propagation.
One should note that bright solitons are formed by modulating the scattering length $a_s (t)$, which controls the magnitude and sign of the interatomic interaction strength. This directly affects the interaction term g(t), with solitons forming when the interactions become attractive (i.e., when $a_s (t)< 0$). This process is independent of the azimuthal index $\ell$ of the L-G beams, as soliton formation relies solely on the modulation of $a_s(t)$. As noted earlier, increasing $\ell$ reduces the number of collisions, slowing the traveling wave and leading to distinct propagation patterns.
\hfill \break
\hfill \break
In figure \ref{fig:Fig5}, we present the soliton dynamics using the second method, which involves introducing a barrier within the initial potential described by:
\begin{equation}
    V_{1D}(z) = U_zz^{2} + V_{b}e^{-z^2/\sigma^2}
\end{equation}
The atom chosen here is $^{87}$Rb. The barrier is released at \( t_b = 80 \, \text{ms} \), meaning that $V_b$ is set to zero at and beyond time $t_b$. Upon releasing this barrier, we observe the formation of 8 dark solitons. This method of generating solitons contrasts with the first method, where the scattering length \( a_s \) is varied over time.
When we increase the parameter $U_z$, tightening the trap, the number of solitons formed remains constant at 8. The stability in the number of solitons can be attributed to the fixed initial conditions set by the barrier potential, which consistently generates the same number of solitons regardless of the trap width.
The same trend, as for the first method, is observed with the increase in $U_z$: the number of collisions between the solitons increases, when $U_z$ increases, from 3 to 4 to 5 for  $U_z = 0.25$,  $U_z = 0.5$, and  $U_z = 0.75$ respectively. A larger $U_z$ corresponds to a tighter trap, which confines the solitons in a smaller spatial region, thus enhancing the frequency of interactions and collisions. This behavior can be explained by the dynamics of solitons in a confined space where their trajectories overlap more frequently, leading to an increased rate of collisions. 
\begin{figure}[h]
    \centering
    \includegraphics[width=1\linewidth,height=5cm]{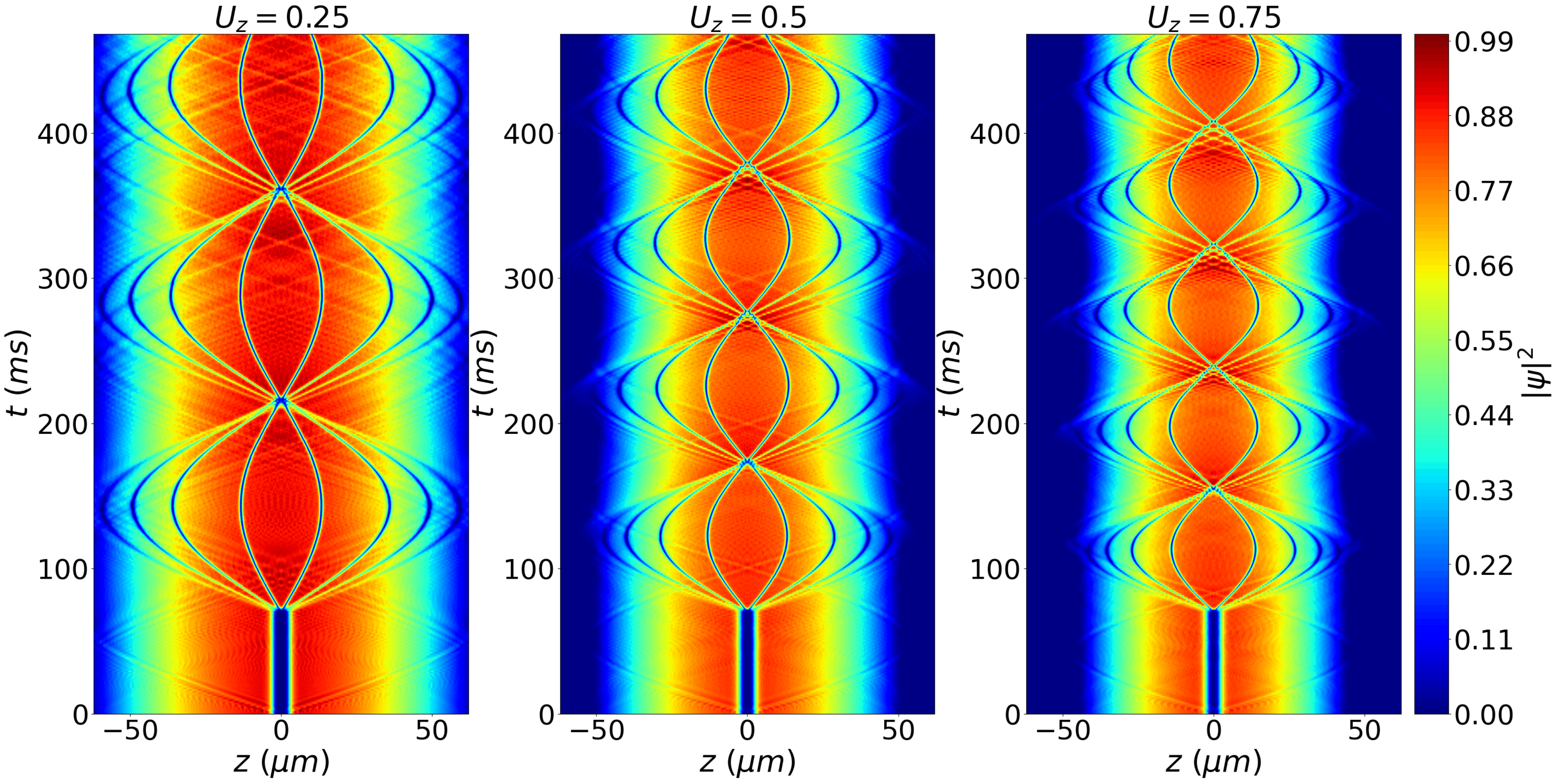}
    \caption{Soliton dynamics generated by adding and removing the barrier for different value of $U_z$ for Rubidium ($^{87}$Rb) atom.}
    \label{fig:Fig5}
\end{figure}
%**********************************
%\paragraph{Anharmonic trap}
%**********************************
\hfill \break
\hfill \break
In the case of an anharmonic trap with $\ell \neq 1$, the dynamics of solitons differ significantly from those in a harmonic trap. Figure \ref{fig:Fig6} illustrates the soliton dynamics for $\ell = 3$ in panel (a) and $\ell = 6$ in panel (b). We apply the same initial conditions as in the second method, where a barrier is introduced and then suddenly removed at $t_b = 80$ ms. For consistency and to facilitate comparison, we took the same atom $^{87}$Rb and we adapted the width $U_z$ of the potential regardless of the index $\ell$. Hence, the generated solitons for each trap will have almost the same space to propagate. The only parameter varying in this scenario is the value of $\ell$. As we increase the value of $\ell$, the trap tends to have a cubic form for higher values (e.g., $\ell = 6$). This change in shape impact the velocity of soliton propagation over time.
The observed differences in number of collisions between solitons highlight the impact of the trap's shape on soliton dynamics. One could observe in figure \ref{fig:Fig6} that the number of collisions decreases when we increase $\ell$ reflecting a decrease in the speed of the travelling wave. This leads to distinct propagation patterns for different trap shapes, underscoring the importance of trap geometry in controlling and manipulating soliton behavior in Bose-Einstein condensates.
\\
In addition, we can observe, in figure \ref{fig:Fig6},  that the soliton dynamics is quite different from the soliton dynamics in the case $\ell=1$. In the harmonic case, the solitons interact solely at $z=0$, which is not the case in an anharmonic trap framework. In the latter case, we can observe the interaction between solitons at different values of $z$. Upon interaction at $z\neq 0$, the solitons manifest evident phase shifts when colliding. This behavior was absent in $\ell=1$ and suggests novelty in interactions between solitons in anharmonic cases.\\
Furthermore, as the values of $\ell$ increase, one can observe broader trajectories of the traveling waves. This phenomenon can be explain by the solitons reflexion on the edges of the BEC which, as $\ell$ increases, tends to a more cubic form (less rounder faces).
\begin{figure}[h]
    \centering
    \includegraphics[width=0.8\linewidth, height = 5 cm]{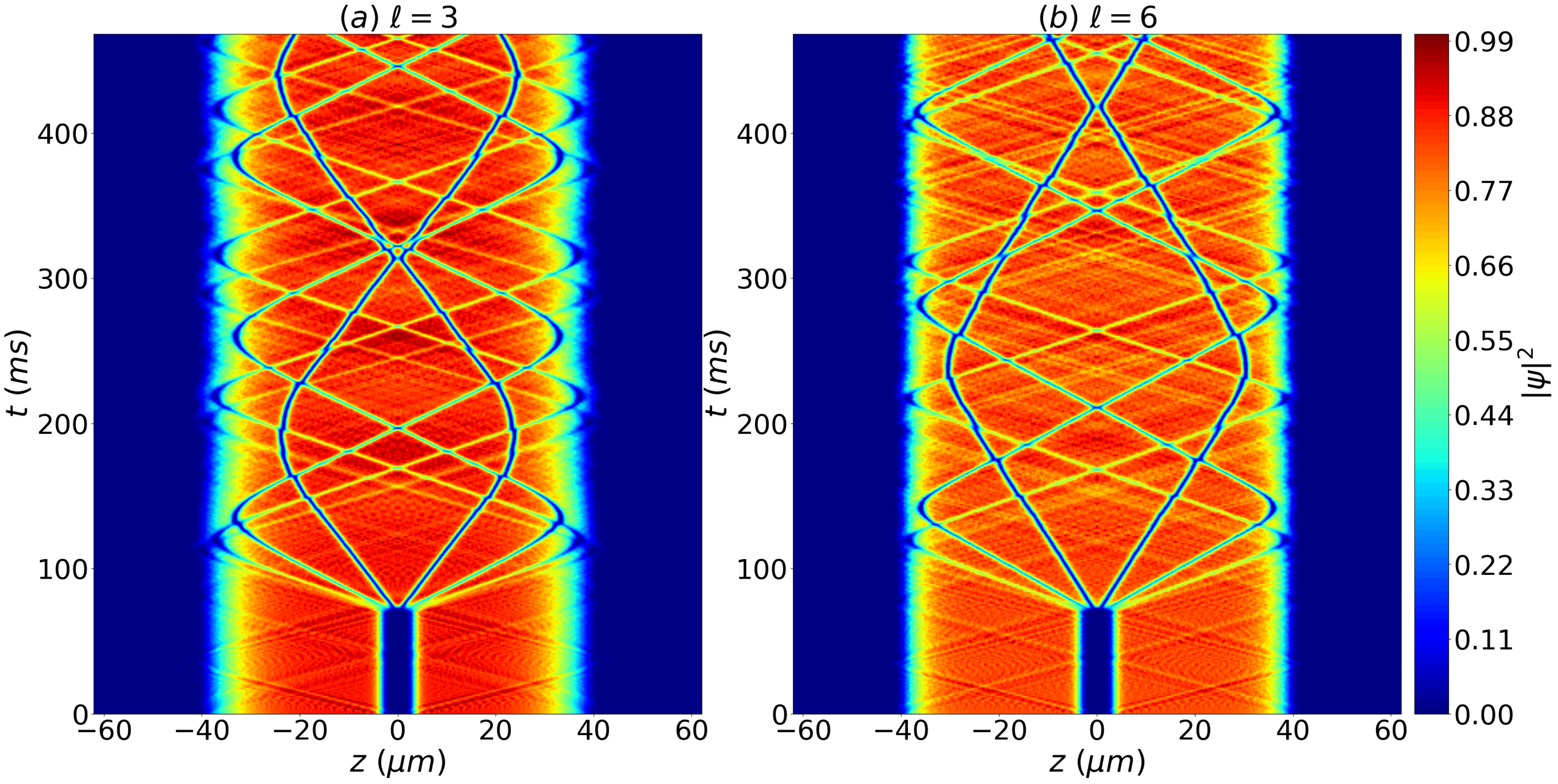}
    \caption{Soliton dynamics generated by adding and removing the barrier, panel (a) for $\ell =3$, and panel (b) for $\ell = 6$, for Rubidium ($^{87}$Rb) atom.}
    \label{fig:Fig6}
\end{figure}
%*****************************************
\subsubsection{Effect of the mass}
\begin{figure}[h]
    \centering
    \includegraphics[width=1\linewidth, height = 4cm]{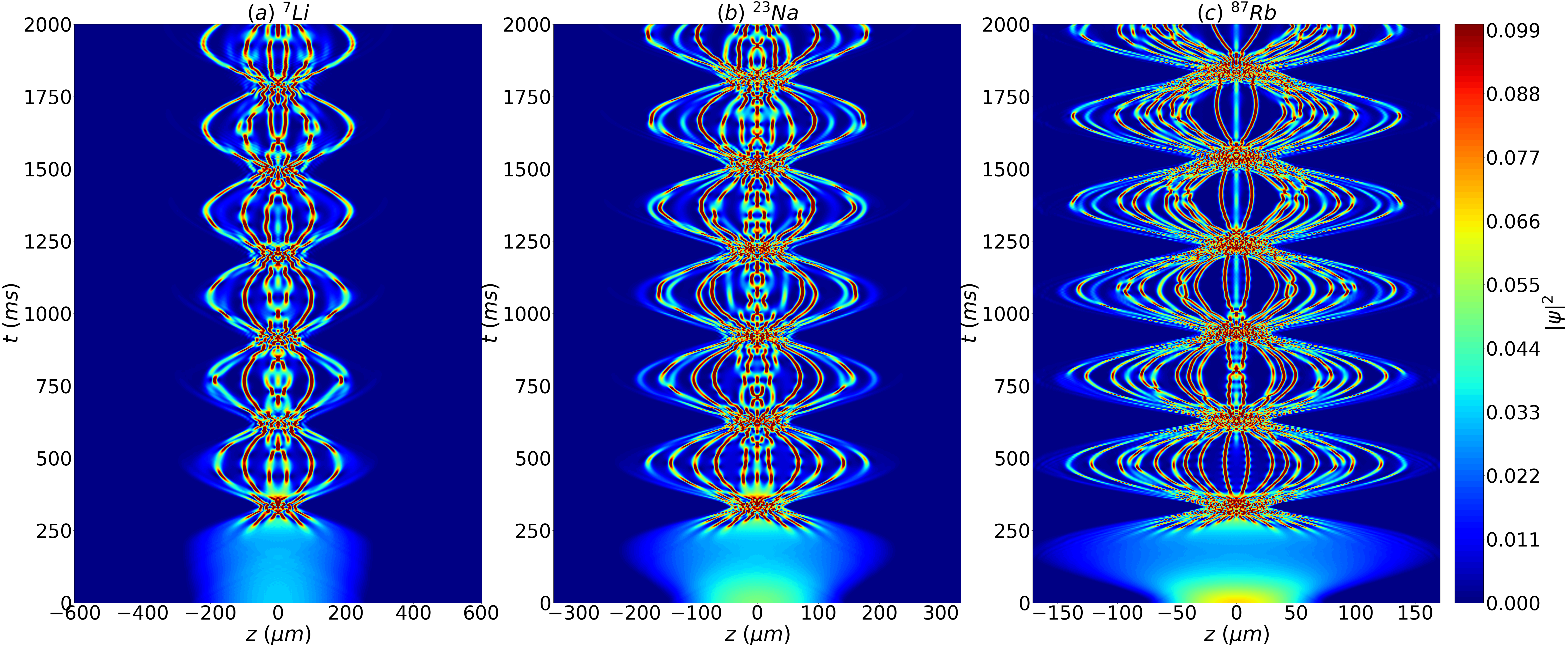}
    \caption{Solitons generation for different atoms, panel (a) for Lithium ($^7$Li) atom, panel (b) for Sodium ($^{23}$Na) atom, and panel (c) for Rubidium ($^{87}$Rb) atom.}
    \label{fig:Fig7}
\end{figure}

%*****************************************
In this subsection, we explore the impact of the atomic mass on the creation and dynamics of solitons in Bose-Einstein condensates. The mass of the atoms plays a crucial role in determining the interaction strength within the condensate, which in turn influences the formation and behavior of solitons. We recall that the interaction strength \( g = \frac{4\pi\hbar^2a_s}{m} \). From this relationship, it is evident that the interaction strength \( g \) is inversely proportional to the mass \( m \). To investigate the effect of mass on soliton formation, we consider three different atomic species: Rubidium ($^{87}$Rb), Sodium ($^{23}$Na) and Lithium ($^7$Li). These species cover a wide range of atomic masses, providing an understanding of how mass influences soliton dynamics.
Figure \ref{fig:Fig7} illustrates the number of solitons formed for each atomic species, we adopt here the first method of changing the scattering length $a_s$ over time. We set the value of $U_{z} = 0.5$ for comparison purposes. As the atomic mass increases from Lithium to Sodium to Rubidium, the number of solitons also increases from 4, to 8 to 16 respectively. This trend can be explained by the inverse relationship between mass and interaction strength. For heavier atoms, the interaction strength is reduced, leading to a higher tendency for soliton formation due to weaker repulsive interactions.
Specifically, for $^7$Li atom with a small mass, the interaction strength \( g \) is relatively high, resulting in fewer solitons. For $^{23}$Na atom with an intermediate mass value, the interaction strength decreases, allowing more solitons to form. Finally, for $^{87}$Rb atom with a quite high mass value, the interaction strength is the lowest among the three, leading to the formation of the highest number of solitons.

We should mention that the atomic mass does influence the terms related to kinetic energy and trap potentials in GP equation, the formation and dynamics of solitons in BECs are primarily driven by the non-linear term, which originates from the interaction strength g. Solitons are generated when the non-linear term, which represents the interatomic interactions, balances the dispersive effects from the kinetic energy term. This balance is what allows the soliton to maintain its shape during propagation. The interaction term, which is proportional to the scattering length and inversely proportional to the atomic mass, plays a critical role in this balance. The change in g through $a_s$, which depends on the mass, more strongly affects the non-linear interaction term and hence dominates the dynamics of solitons. In addition, since solitons are relatively localized structures, their dynamics are more sensitive to the strength of interactions than to the kinetic term. The trapping potential also plays a role, but typically provides a slower mass-dependent confinement effect, which influences the overall shape and oscillations of the condensate but not the localized soliton dynamics as directly as the interaction term does.
%****************************************
\subsection{SOLITON DYNAMICS IN 2D}
\begin{figure}[h]
    \centering
    \includegraphics[width=0.8\linewidth]{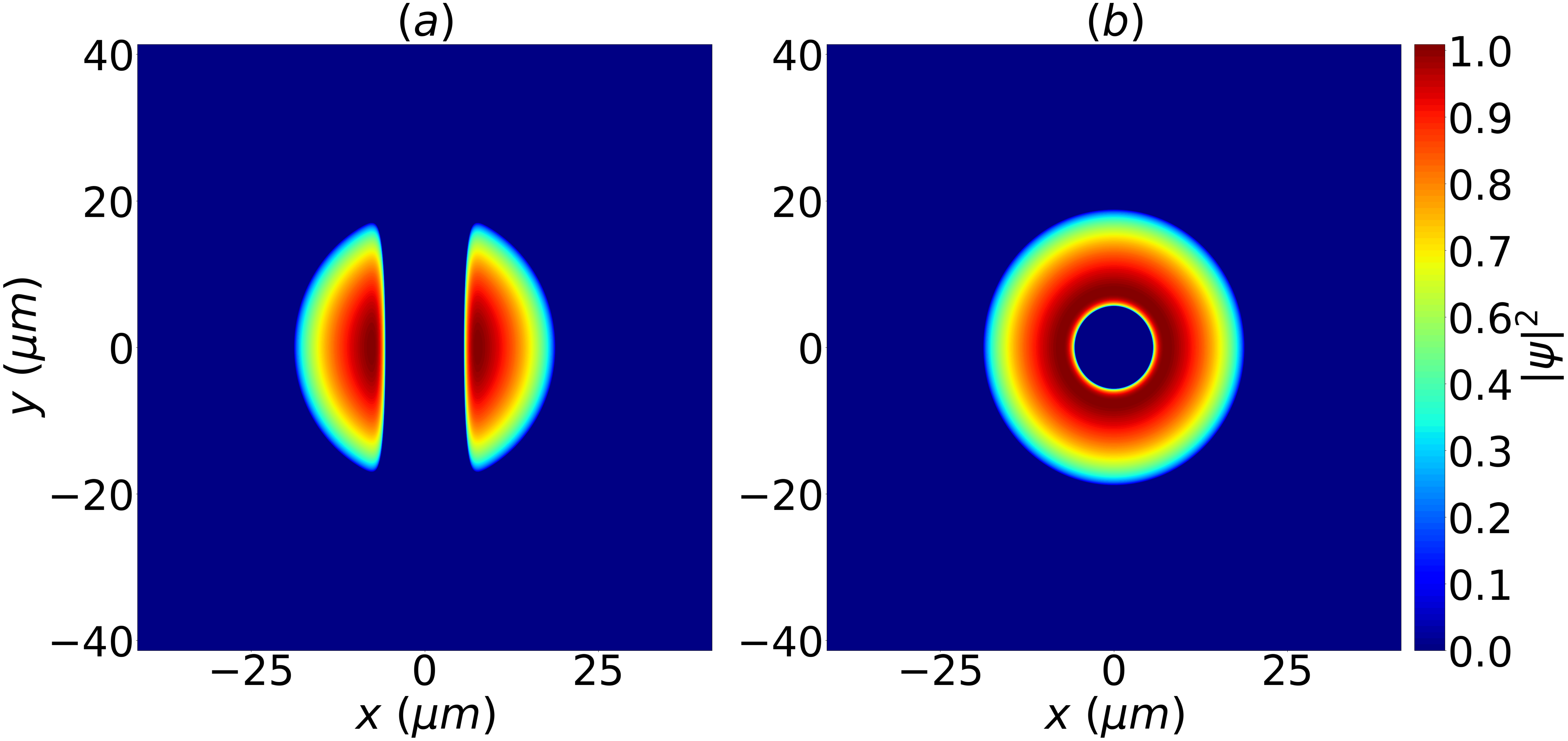}
    \caption{Initial barrier introduced to split the BEC into two symmetric parts (panel (a)). Hole ring introduced in the center of the BEC along $x$ and $y$ axes (panel (b)).}
    \label{fig:Fig8}
\end{figure}

\begin{figure}
    \centering
    \includegraphics[width=1\linewidth, height = 7 cm]{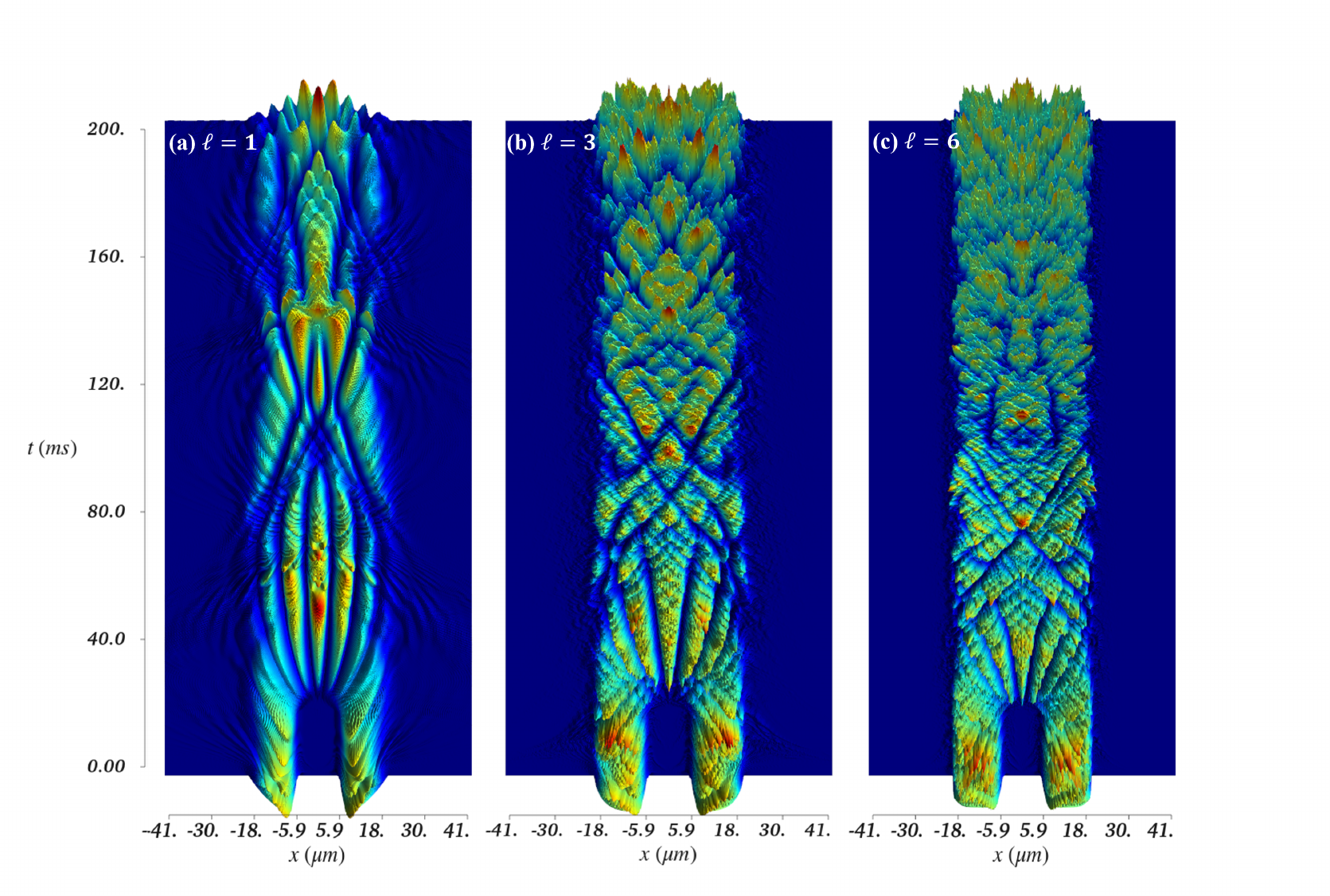}
    \caption{Soliton Dynamics in 2D-BEC following a removal of a barrier along $x-$axis, panel (a) for $\ell = 1$, panel (b) for $\ell = 3$ and panel (c) for $\ell = 6$.}
    \label{fig:Fig9}
\end{figure}

%****************************************
%************************************
In the 2D configuration, we initially use the second method to generate the solitons by introducing a barrier only along the $x$-axis, which is then suddenly removed at $t_b = 20$ ms. The potential used in this method is defined as $V_{2D}(\rho) = U_{\rho}\rho^{2\ell} + V_{b}e^{-x^2/\sigma_{x}^2}$. We take, as an example,  $V_b = 326.6$ nk and $\sigma_x = 3.6 \ \mu$m.  The initial BEC wave-function is displayed, in Thomas-Fermi approximation, in panel (a) of figure \ref{fig:Fig8}.
Note that if the barrier were introduced along the $y$-axis instead of the $x$-axis, the results would remain consistent with those observed for the barrier along the $x$-axis. This consistency arises from a rotational invariance of the system, which ensures that the dynamics are independent of the barrier's orientation.
To visualize the dynamics of the solitons, we plot the density distribution of the BEC over time along the $x$ axis at $y = 0\ \mu$m, as shown in figure \ref{fig:Fig9}. Initially, the BEC exhibits a Gaussian density profile with a distinct dip around $x = 0\ \mu$m, indicating the presence of the barrier along the $x-$axis. Upon releasing the barrier at $t_b$, we observe the emergence of dark solitons, formed as a result of the barrier removal. As time advances, these solitons become increasingly well-defined and distinct, showcasing their typical characteristics.  
One could notice that the dynamics of these solitons vary depending on the value of the index $\ell$ in figure \ref{fig:Fig9} (panel (a) for $\ell = 1$, panel (b) for $\ell = 3$ and panel (c) for $\ell = 6$). This variation in soliton's behavior highlights the significant influence of the trap's geometry on the soliton formation and dynamics, when we follow the same method for solitons generation. The observed patterns underscore the complex interplay between the trap's geometry and the resulting soliton dynamics, suggesting that careful tuning of the trap parameters could be used to control soliton properties and behaviors in experimental settings.

In our exploration of the 2D scenario, we also investigated the introduction and subsequent sudden removal of a Gaussian hole, defined by the potential: $V_{2D}(\rho) = U_{\rho}\rho^{2\ell} + V_{h}e^{-\rho^2/\sigma_{\rho}^2}$. This setup creates a central dark hole within the BEC, as illustrated in panel (b) of figure \ref{fig:Fig8}. The BEC then takes the form of a ring. When the Gaussian hole is removed at the same time $t_b$, the resulting patterns from this configuration differ significantly from those observed in figure \ref{fig:Fig9}, where a barrier was used along the $x$-axis only.
\begin{figure}
    \centering
    \includegraphics[width=1\linewidth, height = 7 cm]{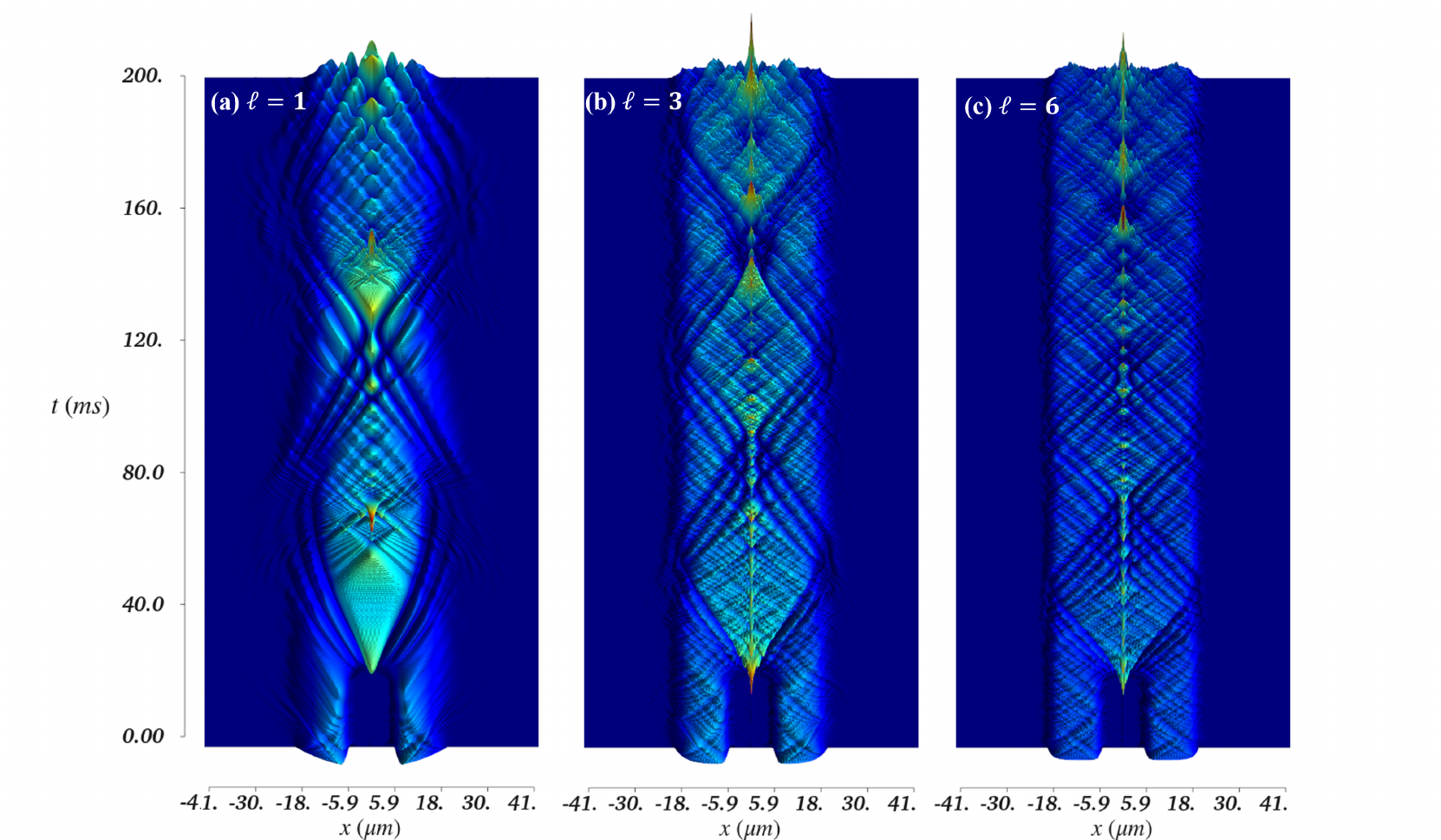}
    \caption{Soliton Dynamics in 2D-BEC following a removal of a gaussian hole, panel (a) for $\ell = 1$, panel (b) for $\ell = 3$ and panel (c) for $\ell = 6$.}
    \label{fig:Fig10}
\end{figure}
In the case of the Gaussian hole, we observe that the dark solitons tend to form near the edges of the BEC and follow straighter trajectories. These solitons collide near the center and continue their paths, as seen in panels (b) and (c) of figure \ref{fig:Fig10} for $\ell = 3$ (panel (b)) and $\ell = 6$ (panel (c)), respectively. The solitons oscillate consistently over time, maintaining their coherence and direction. 

In contrast, in the scenario with the barrier along the $x-$axis, the solitons exhibit different behavior. As shown in figure \ref{fig:Fig9}, particularly in panels (b) for $\ell = 3$ and (c) for $\ell = 6$, the solitons appear to lose their trajectories around $t =100$ ms. This loss of coherence could be attributed to the increased complexity of interactions in the absence of the barrier, leading to more chaotic dynamics compared to the Gaussian hole scenario. The difference in soliton behavior underscores the significant impact of the potential landscape on the dynamics and stability of solitons in BECs.

The 2D visualizations highlight the interaction patterns among solitons, which differ somehow from the 1D case. The ability of solitons to move in two dimensions leads to a different set of dynamics. The geometry of the trap, dictated by the LG beam azimuthal index $\ell$, directly influences these dynamics. As $\ell$ increases, solitons exhibit distinct behaviors, such as increased frequency of collisions and altered trajectories, as they are more constrained in their movements.
In general, visualization of solitons in 1D and 2D provides interesting insights into their formation and dynamics within different trapping potentials. The shape of the trap, particularly the degree of anharmonicity and the control provided by LG beams, determine the behavior and interaction of solitons, with implications for experimental realizations and potential applications in quantum technologies.

This study is primarily theoretical, the proposed phenomena, particularly the generation of bright solitons via scattering length modulation, would be experimentally feasible. Previous experiments involving BECs in similar trap configurations have been proposed and successfully executed using current experimental techniques. In particular, Feshbach resonances have been effectively utilized to modulate the scattering length in highly elongated 1D traps. This controlled modulation, transitioning the interactions from repulsive to attractive, has been shown to lead to soliton formation and propagation, as demonstrated in experiments with lithium condensates \cite{Strecker2002, khaykovich2002, marchant2013}.
One potential challenge in the experimental realization of solitons is maintaining their stability during periodic collapses that can occur at high densities, especially when attractive interactions are strongly induced. At such densities, the GP equation, which assumes mean-field theory and considers only two-body interactions, may become insufficient. The emergence of higher-order effects, such as three-body collisions, can destabilize the condensate, leading to rapid loss of atoms and preventing stable soliton formation \cite{Cornish2006}. However, by carefully controlling the magnitude and duration of the negative scattering length, these issues can be mitigated. Avoiding excessively negative values of the scattering length and maintaining lower atomic densities can help keep the condensate in a metastable state long enough to observe bright soliton formation.
Furthermore, the use 1D geometries, where radial confinement is strong and the dynamics occur primarily along the axial direction, helps suppress higher-order effects, making the GP equation a more reliable approximation for soliton dynamics \cite{carr2002, Nguyen2014}. 1D trapping has been successfully implemented in multiple experiments, providing a method to stabilize solitons while minimizing the destabilizing effects of three-body collisions. 
By addressing potential challenges such as higher-order scattering effects and periodic collapse through careful experimental design, this work provides a sort of framework for future experimental investigations into soliton behavior in Bose-Einstein condensates.
%************************************
%\subsubsection{Anharmonic trap}
%***********************************
%*******************************
\section{CONCLUSION}
%*******************************
In this study, we have explored the generation and dynamics of solitons in BECs within shaped potentials created by two crossed LG beams. Utilizing the distinctive properties of these beams, we have studied different techniques to facilitate solitons formation in both harmonic and anharmonic traps by only changing the azimuthal index $\ell$. Our theoretical framework was built on the derivation of the Gross-Pitaevskii equation (GPE) under specific conditions, followed by numerical simulations to investigate the soliton dynamics. We employed two primary methods for generating solitons: the first method demonstrated the effectiveness of changing $a_s$ from a positive to a negative value to induce attractive interactions, leading to bright solitons formation. The second method involved using a potential with an initial barrier, which upon removal, allowed the creation of dark solitons. 
Our results indicate that the shape and parameters of the trap significantly affect the number and dynamics of solitons. In particular, we observed that anharmonic traps with higher azimuthal indices $\ell$ resulted in fewer solitons due to tighter confinement, which also influenced their trajectories and interactions. Furthermore, the mass of the atomic species in the BEC was shown to impact the soliton formation, with heavier atoms leading to an increased number of solitons.
The ability to control the soliton dynamics in BECs holds significant promise for a wide range of applications, including advances in quantum technologies and precision sensing.

There are several exciting avenues for future research. Exploring the impact of varying the parameters of the LG beams, the radial index $p$ as well and the trapping potentials could provide further insights into optimizing soliton generation and control. Furthermore, we could also investigate the creation of vortices in BEC \cite{Brachmann11}. The potential applications of controlled soliton dynamics in quantum technologies and precision sensing highlight the importance of continued research in this area. By deepening our understanding of solitons in BECs, we can contribute to the development of innovative technologies and enhance our ability to manipulate and control quantum systems.
\hfill \break

\textbf{Data availability statement}
All data supporting the findings of this study are included in the article.

\begin{acknowledgments}
This project is part of the multidisciplinary program supported by an internal fund from the ECE-Paris Graduate School of Engineering, a member of Omnes Education.
\end{acknowledgments}

\bibliography{Bibliography}% Produces the bibliography via BibTeX.

\end{document}